\documentclass[11pt,twoside]{article}
\usepackage{asp2010}

\usepackage{graphicx} 
\usepackage{url}

\resetcounters

\begin{document}

\title{Teaching Astronomy with an Inquiry Activity on Stellar Populations}
\author{Marc Rafelski$^1$, Michael Foley$^2$, Genevieve J. Graves$^3$, Katherine A. Kretke$^4$, Elisabeth Mills$^5$, Michael Nassir$^6$, and Shannon Patel$^4$
\affil{$^1$Department of Physics and Center for Astrophysics and Space Sciences, University of California, San Diego, La Jolla, CA 92093, USA}
\affil{$^2$Hawaii Natural Energy Institute, School of Ocean And Earth Science And Technology, University of Hawaii at Manoa, Honolulu, HI 96822, USA}
\affil{$^3$Department of Astronomy, University of California, Berkeley, CA 94720, USA}
\affil{$^4$Department of Astronomy and Astrophysics, University of California, Santa Cruz, CA 95064, USA}
\affil{$^5$Division of Astronomy and Astrophysics, University of California, Los Angeles, CA 90095, USA}
\affil{$^6$Department of Physics and Astronomy, University of Hawaii at Manoa, Honolulu, HI 96822, USA}}

\begin{abstract}
We describe a new inquiry design aimed at teaching advanced high-school to senior college students the basics of stellar populations. The inquiry is designed to have students come up with their own version of the Hertzsprung-Russell diagram as a tool to understand how stars evolve based on their color, mass, and luminosity. The inquiry makes use of pictures and spectra of stars,  which the students analyze and interpret to answer the questions they come up with at the beginning. The students undergo a similar experience to real astronomers, using the same tools and methods to figure out the phenomena they are trying to understand. Specifically, they use images and spectra of stars, and organize the data via tables and plots to find trends that will then enable them to answer their questions. The inquiry also includes a ``thinking tool'' to help connect the trends students observe to the larger picture of stellar evolution. We include a description of the goals of the inquiry, the activity description, the motivations and thoughts that went into the design of the inquiry, and reflections on how the inquiry activity worked in practice. 
\end{abstract}

\section{Introduction}

Observational astronomy is different from many sciences because astronomers lack the ability to run laboratory experiments to learn how the universe works.  There are two main impediments astronomers have to deal with: first, the objects of study are too massive to create or fit in a laboratory, and second, the objects of study often change over time-scales of millions of years. For instance, astronomers cannot learn how mass affects the evolution of stars by creating two stars of different masses in the lab and watching them evolve. Nor can they learn about galaxy interactions by throwing different types of galaxies at each other. Instead, astronomers generally have to rely on observations of objects in their present form, and link objects with similar characteristics at different stages of their evolution together to learn how they change over time.

Astronomers generally have three types of tools at their disposal: they can take pictures, spectra, and repeat observations over time. The last tool only works for objects that change on short time scales, which is a limited but important subset of objects in astronomy, such as variable stars, supernovae, and pulsars. The other two tools are the largest workhorses in astronomy, and can be combined in different ways. For instance, images taken covering different wavelengths can be combined to create color images that we show the public. 

In general, inquiry activities create learner-centered environments that can increase student interest while they learn content and scientific process skills. They also increase students' confidence in their ability to do science and give them an appreciation of what it means to do science \citep{chinn02, hpl}. In this inquiry, we set out to not only provide students this learning environment and increased science appreciation, but also specifically provide them an insight into how astronomers do science. We simulate the experience of an astronomer by giving the students the same materials that astronomers use, namely pictures and spectra. We also give them some quantitative properties that would normally be derived from these materials. They use these tools to learn about stellar populations and a large number of attitudinal and process goals outlined in \S3. Those readers interested in merely a description of the inquiry activity may skip to \S4.

The stellar populations inquiry was designed as part of the Professional Development Workshop \citep{PDP}, and has been taught twice at the time of this proceeding, once at a ``short course'' taught at the University of California, Santa Cruz, in 2007, and once as part of the Po`okela program in Maui, Hawaii in 2008. The materials used in this inquiry were all printed out on paper, and therefore this inquiry could easily be modified to work in large introductory astronomy classes. 
The materials and color figures can be found on a webpage for this inquiry at: \\
 \url{http://stellarpopulations.pbworks.com/}

\section{Audience and Expectations}

This inquiry is designed to work with a large variety of students, ranging from advanced high school students to senior college students. It is meant for students who are not pursuing degrees in science, or are at an early stage in their careers. The inquiry was initially designed for entering college students, and later modified to work with high school students. While the inquiry activity works for different audiences, the expectations of the students should correspond to their grade level. For instance, when teaching college students we would expect them to make graphs, while a more experienced college group might make logarithmic plots. On the other hand,  when teaching high school students we would accept tables of data, with the more experienced students making plots. 

This inquiry was designed to follow the inquiry called Color, Light, and Spectra (CLS) designed by Matthew Barczys, Seth Hornstein, and Lynne Raschke.  If the stellar populations inquiry is not taught in conjunction with the CLS inquiry, the teacher will need to cover the content from CLS beforehand.  
Specifically, knowledge that color filters transmit different wavelengths of light, and an understanding of blackbody radiation sources, along with the relationship between blackbody peak wavelength and temperature, are needed. Additionally, CLS helps the students get familiar with what a spectrum is and the difference between continuous light and discrete lines. We also do not recommend this being the students' first inquiry activity, as it is more difficult than in other inquiry activities for students to ask good questions from the intriguing phenomena presented (called starters) that match the content goals, and the materials somewhat limit the range of questions the students can investigate. Both times the stellar populations inquiry was taught, the students undertook the CLS inquiry first. 

Lastly, the students' prior knowledge can vary drastically coming into this inquiry, with some students being comfortable with logarithms and graphing, while others are not. Also, some students are quite familiar with astronomical terms and ideas, while others have no knowledge at all on the subject. Depending on your audience, more or less prior knowledge needs to be conveyed to the students. It is also important to note that while much of the prior knowledge is useful, sometimes the students' preconceptions are wrong and can hinder the students' advances \citep{McDermott91, redish94,hpl}. It is up to the facilitators of the inquiry to identify these misconceptions and address them.

\section{Goals for learners}

We approached the design of this inquiry using backward design \citep{wiggins05} in which scientific process, attitudinal, and content goals are defined first.  The inquiry activity is then designed to convey these goals. In that spirit, we outline the goals here before describing the inquiry. Given the venues this inquiry was designed for, the inquiry process and attitudinal goals were more important than the content goals. More content could easily be incorporated with the materials given. Additionally, the materials could be used more than once to cover different content goals in an introductory astronomy course. For instance, changing the starters to pictures of galaxies would yield very different questions to investigate, and images and spectra of stars would help answer such questions. Since the students would already be familiar with the materials, secondary inquiries might take substantially less time. While our goals were specific for the venues the inquiry would be taught in, we found that our skills and attitudinal goals are especially quite well aligned with goals for Astronomy 101 classes determined by a set of national NSF workshops \citep{astro101}. The goals are all related to each other, but we separate them to facilitate the discussion. 

\subsection{Inquiry Process Goals}

$\bullet$ Interpreting data (recognizing patterns  and trends) \\
$\bullet$ Selecting relevant data from too much data \\
$\bullet$ Understanding goals of the scientific process \\
$\bullet$ Evidence based critical thinking \\ 
$\bullet$ Interpreting and applying new data with recently acquired content \\

\noindent The process goals are the most important part of this inquiry. We want the students to learn to interpret data the way scientists do. We accomplish this by requiring the students to recognize patterns and trends from a body of data that is much larger than what they need. Scientists often have a lot of data, and make many plots to see if they can find trends that make sense, and then apply those trends to the scientific questions they are trying to answer. The students doing this inquiry are meant to go through a similar process. By experiencing the scientific process, we hope that they gain an understanding of what the process is beyond just some words in a textbook. Moreover, we want the students to spend time thinking critically to answer questions based on the evidence they have. It is useful to remind the students of what you want them to learn, including this problem solving skill. The students' metacognition will help them improve and retain these important skills learned throughout the inquiry \citep{Bransford86,Flavell79, Gourgey98}. Lastly, we want the students to take what they have just learned, and apply it to learn something new. This is a large part of the reason that the CLS inquiry was made an integral component of teaching this inquiry in the past. 

\subsection{Attitudinal Goals}

$\bullet$ Open mindedness and willingness to change theory in light of evidence \\
$\bullet$ Appreciation for astronomy \\ 
$\bullet$ Connect inquiry with daily lives \\
$\bullet$ Excitement about STEM (Science, Technology, Engineering, and Mathematics) \\
$\bullet$ Community building \\
$\bullet$ Being confident in participating in scientific discourse \\

\noindent Similar to the process goals discussed above, we want the students to have an open mind and form theories that they are willing to change in light of the evidence from their investigations. We also want the students to come out appreciating astronomy and being excited about science. Furthermore, we want the students to realize how astronomy is part of their daily lives, and build a community in the group of students. Lastly, in one of our venues the students moved on to do internships afterwards, and we wanted the students to be confident in participating in scientific discourse at those internships. 

\subsection{Content Goals}

$\bullet$ Blue stars are bright and hot, red stars are dim and cool \\
$\bullet$ Blue stars have shorter lives than red stars \\
$\bullet$ Stars in clusters have the same age \\
$\bullet$ Red clusters are older than blue clusters \\
$\bullet$ Stars form from gas clouds \\

\noindent The content goals are straightforward, constituting the basics of stellar populations. 

\section{Activity Description}

The activity has many components, which we outline in Table 1 below. We split the inquiry activity into two days because the first couple of experiences a student has with the inquiry process can be frustrating due to the open-ended and self-directed nature of the activity. Splitting up the activity gives the students a break, while giving them time for self reflection. Each of the major components is outlined in its own subsection below. 

\begin{table}[!ht]
\caption{Stellar Populations Inquiry Activity Outline}
\smallskip
\begin{center}
{\small
\begin{tabular}{lclc}
Activity Description Day 1 & Time & Activity Description Day 2 & Time\\
\noalign{\smallskip}
\tableline 
\noalign{\smallskip}
Introduction & 10 min & Personal investigations & 30 min \\ 
Starters & 10 min & Thinking tool & 20 min\\ 
Group question session & 20 min & Personal investigations & 50 min\\ 
Selection of  questions & 10 min & Make posters & 20 min\\ 
Personal investigations & 60 min & Present posters & 20 min\\ 
 & & Synthesis and conclusion & 20 min\\ 
  \noalign{\smallskip}
\tableline 
\noalign{\smallskip}
 Total time & 110 min& Total time & 160 min
\end{tabular}
}
\end{center}
\end{table}

\subsection{Introduction}
While in general the introductions to inquiries may be straightforward, in this inquiry it is very important to discuss a few topics with the students. First, it is essential to explain that in astronomy we cannot touch our science, and so we are left with tools like spectra and images to gain new knowledge. Second, this is an important time to mention that they will be using problem solving skills that are important not only in astronomy, but in other fields of study. It is good to make them conscious of their use of critical thinking skills.  This metacognition of their learning and use of problem solving will not only help them perform better at the task, but it will also help retention of those skills \citep{Bransford86,Flavell79, Gourgey98,  lin99, lin01}. Third, this is a great time to remind them of the content they learned in CLS (such as colors of blackbodies, and spectra), and again, make them aware that they will be using this new knowledge to gain more knowledge. Lastly, depending on the level of the students, it might help to suggest that they figure out ways to organize the data, and perhaps mention that scientists make many plots. Explain why graphs and tables are important and how they help to organize thinking, and make it easier to look for physical relationships in the data. If an educator implementing this inquiry has other goals than those we described above, this would be a good place to address them.

\subsection{Starters}
The starters are pictures of stars in different astronomical situations which are meant to get the students to ask lots of questions, of which they will later choose one to investigate. The students should look at one of these pictures at a time, and write down their questions on large sentence strips (large strips of posterboard paper that can be posted on the wall and seen from a distance). If it is the students' first time coming up with questions, the facilitators will probably have to model what a good question looks like. Students have a hard time coming up with questions, so we generally encourage them to write as many as possible. The more times the students get to practice asking questions, the better they get at it. 

The starter pictures are carefully chosen to show a few different phenomena. First, they show stars of a myriad of colors and brightnesses. Second, they show star clusters with different colors. Lastly, they show blue young clusters with nebulosity (gas and dust) around them. These characteristics are specifically chosen as they lead to questions that address our content goals, while being investigable with the given materials. We minimized the number of diffraction spikes and rings caused by the CCDs, and other distractions in the images. Also, we avoided pictures where it was difficult to discern boundaries of the stars and star clusters. The starters are given in Table 2, while alternative possible starters are on the stellar populations website.

\begin{table}[ht]
\caption{Images for Starters}
\smallskip
\begin{center}
{\small
\begin{tabular}{lcl}
\tableline
\noalign{\smallskip}
Name & URL & Description \\
\noalign{\smallskip}
\tableline 
\noalign{\smallskip}
M35 \& NGC 2158 & 1&  CFHT image with a red and a blue star cluster \\
M67 &2 & CFHT image of a blue and white star cluster \\ 
NGC 6093 & 3 & Hubble image of a red star cluster \\ 
Omega Cen &4 & Red star cluster \\ 
M16 &5 & Star forming region with gas and dust \\ 
NGC 3606 &6 & Star forming region with gas and dust \\ 
\tableline
\multicolumn{3}{l} {\tiny[1] $\tt http://www.cfht.hawaii.edu/HawaiianStarlight/AIOM/English/2003/Images/Dec-Image2002-CFHT-Coelum.jpg$}\\
\multicolumn{3}{l} {\tiny[2] $\tt http://www.cfht.hawaii.edu/HawaiianStarlight/AIOM/English/2005/Images/Dec-Image2004-CFHT-Coelum.jpg$}\\
\multicolumn{3}{l} {\tiny[3] $\tt http://heritage.stsci.edu/1999/26/images/9926a.jpg$}\\
\multicolumn{3}{l} {\tiny[4] $\tt http://antwrp.gsfc.nasa.gov/apod/image/0704/OMC-Ver1.jpg$}\\
\multicolumn{3}{l} {\tiny[5] $\tt http://astrim.free.fr/gallery/M16C66.jpg$}\\
\multicolumn{3}{l} {\tiny[6] $\tt http://www.nasa.gov/images/content/191853main\_image\_feature\_929\_full.jpg$}\\
\tableline
\end{tabular}
}
\end{center}
\end{table}

\subsection{Group Question Session}

We find that the question session is fastest if done as a group, although other methods are also effective if more time is available (see other inquiry activities in this volume). If doing a group question session, show each of the pictures on a projector, and have everyone work on writing questions of the same picture at the same time. While students are coming up with questions for the current picture, the facilitators can get the completed question strips from the previous picture and start sorting them by categories to facilitate the students selecting questions (see \S4.4). 
For our content goals, the following categories did a good job: colors of stars, colors of clusters, brightness, origins, Gas/Nebula, and QWWNDWATT.  Here are some example questions: \\

\noindent $\bullet$ Why are stars different colors? \\
$\bullet$ Why are some stars brighter than other stars? \\
$\bullet$ Why are some clusters red, and others blue? \\
$\bullet$ Why do some of the clusters have more red stars than blue stars? \\
$\bullet$ Why do some clusters have gas around them while others do not? \\

Given the nature of the starter pictures, there are many questions asked that do not coincide with our content goals, or are difficult to answer with the materials at hand. We call these questions QWWNDWATTs (questions we will not deal with at this time), and make it clear to the students before they start writing questions that a lot of their questions will fall into this category. In addition, some of the questions are easy to answer quickly, and fall outside of the content goals. Many of these could almost be answered with yes and no answers. We find it useful to just quickly answer these questions with the students at the beginning of the next phase (selection of questions). However, it is important here to make sure not to give away information for the students' own investigations.

\subsection{Selection of Questions (Gallery Walk)}

Once the questions are all sorted and put up on the wall by category, the students walk around reading the questions. They get to pick one question, which they will answer in groups of two. We find that pairs work better than groups of three in this inquiry, since it is easy for one person to get disconnected from the investigation in groups larger than 2. The facilitators help the students get into these groups, and if needed, help rephrase the questions to be well worded. Specifically, a facilitator can help shift a question answerable by a `yes' or a `no' into a question that might use `why', making it a more suitable question. For example, the question `Are some stars brighter than other stars?' could be rephrased to `Why are some stars brighter than others?'.

\subsection{Personal Investigations}

\begin{figure}[htbp]
  \centering
   \includegraphics[scale=0.75, viewport=30 200 530 370]{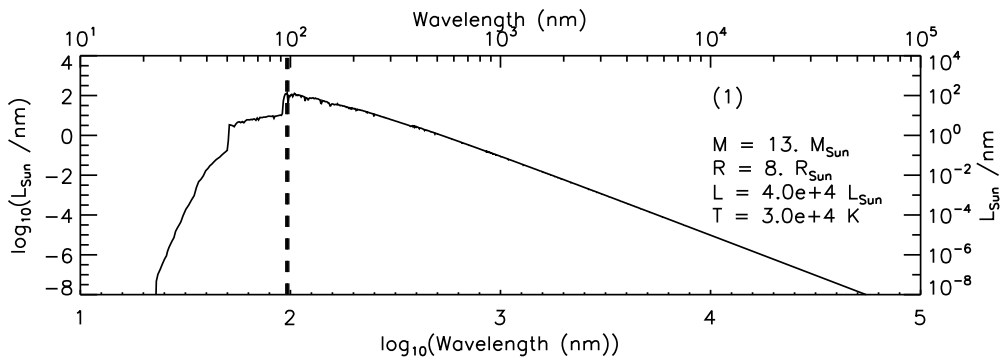} 
   \caption{Example of a spectrum given to the students. The students get 10 of these, with the number in parentheses corresponding to one of the stars labeled in the Sgr. star cloud.}
   \label{spectra}
\end{figure}

The investigations start with an explanation of the materials. The materials are somewhat complicated, so it is useful to go over them together. The students are given copies of all the starter pictures, so they can look at them as often as they want. The students are also given spectra of 10 stars not ordered in any way. Each spectrum has the mass, radius, luminosity, and temperature written on it for that star. The students will need to be told that for this inquiry they can treat luminosity and brightness as basically the same (although they are different!). In addition, the spectra have the visible spectrum color coded by the wavelength it represents. This could be left out for more advanced students. Figure \ref{spectra} shows an example spectrum. It is important to go over what a spectrum represents, and what the axes mean. 

If it is a more experienced group, then less information can be given to the students, so that they have to figure it out. For instance, the spectra have a line marking the peak wavelength of the blackbody. From this, they could calculate the temperature from Wien's law. When we taught the inquiry, we found that the students were not at this level, and therefore we gave them all the information they needed. 

Along with these spectra, the students also get a picture of the Sagittarius star cloud. We do not tell them this is a cluster, but use it because it has a nice sample of stars of different colors. For each spectrum the students receive, one of the stars in this picture is labeled with the number also labeled on the spectrum. The star identifications are based on color alone. Technically, the images should also correspond with luminosity, but this was not possible. Therefore, we found it easiest to tell the students that the stars could be at different distances from us, so the brightness in the image may not reflect their true brightness, even though in this case this is not true since it is a star cloud.

In addition to the ten stars already mentioned, there is an 11th star that is not given to them. This star is a red giant, and can be used in a myriad of situations. The most common situation is if the students think they are done, and have answered their question early. At this point the facilitator can pull out the the 11th spectrum, and ask the students to make sure their explanation also works with this star. Another possibility is that students have come up with an incorrect answer to their question. In this case, sometimes the red giant can be helpful as well as evidence that it may not fit with their theory. We caution against the use of the 11th star if the students are very confused, as it may just confuse them more. It is a great tool for the facilitators to have if used properly. If students still feel done after the 11th star, facilitators should encourage the students to apply what they learned to answer the other questions, especially the ones relating to the gas around stars (if they did not start with one of those questions).

\begin{figure}[htbp]
  \centering
   \includegraphics[scale=0.39, viewport=30 20 500 350]{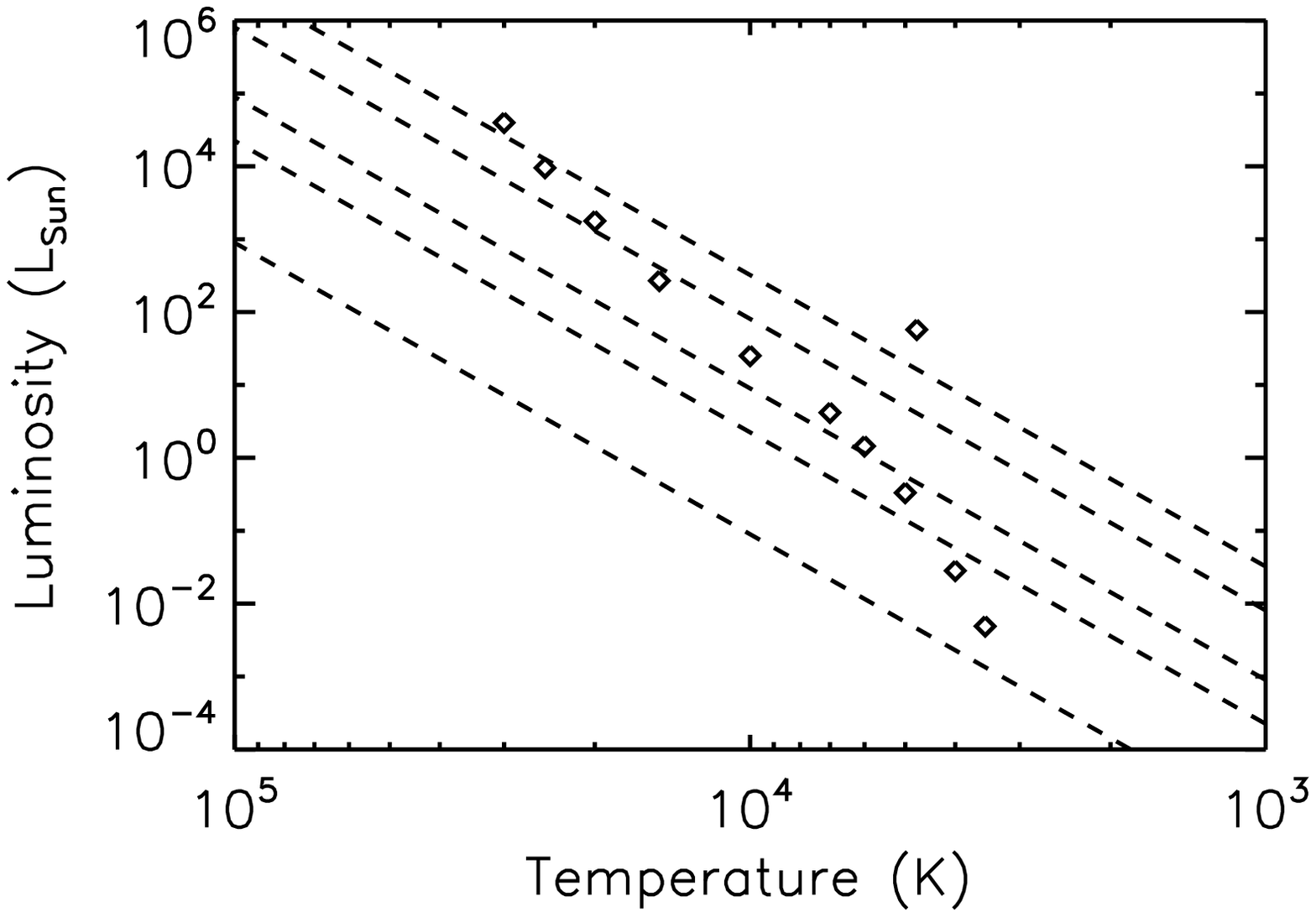} 
    \includegraphics[scale=0.39, viewport=30 20 500 350]{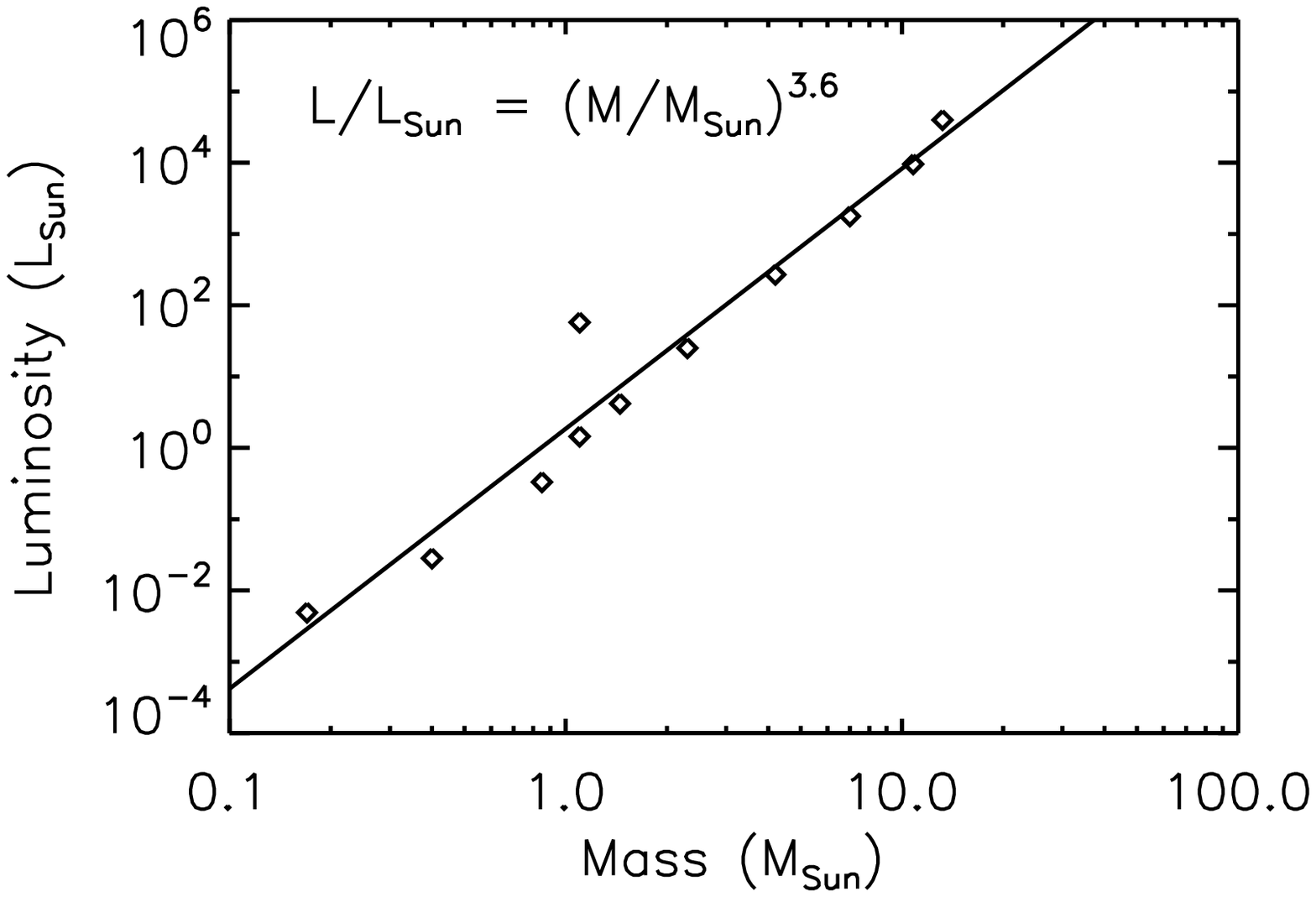} 
   \caption{Examples of plots students may make, although less experienced students (or high school students) will not make log plots like these. The outlier is the 11th star mentioned above, the red giant.}
   \label{exp}
\end{figure}

During the investigation, we are looking for students to organize the data in some way, such as a table or a plot. For more experienced students, we require plots at some point, and many times the students will make these on their own, unprompted by the facilitator. However, if they do not organize the data by the second day, the facilitator needs to encourage them to do this. Once the students have plots similar to Figure \ref{exp}, then these trends will help them answer their question. 

There is one stumbling block the students will encounter, which is how to move from trends of mass, luminosity, radius, temperature, and color to the evolution of stars, and the questions relating to gas. For this to occur, the students need some prompting to think about stellar lifetimes, and the conversion of mass into energy. The first time we taught it, we had the facilitators go over this with the students when they reached the appropriate point. However, we wanted the inquiry to be less facilitator driven, and therefore came up with the thinking tool explained below. The thinking tool should be enough to get the students thinking about stellar evolution, but in some cases the facilitators may still need to help guide the students. It is important to be careful at this step. We want the students to figure out their question on their own in order for them to keep ownership of the material, to help with student motivation and retention of the learned information \citep{hpl}.
All the materials are available at the stellar populations website\footnote{ \url{http://stellarpopulations.pbworks.com/}}.

\subsection{Thinking tool}

The purpose of the thinking tool is to give students a way to think about stellar lifetimes, characteristics of stars, and the idea of converting mass into energy. The thinking tool does this by making an analogy between stars and fuel burning in two containers of different sizes.  In our setup, two metal dishes of drastically different diameters were filled with liquid fuel and ignited. We used a one foot diameter dog dish and a half inch diameter stainless steel cap, and small amount of tiki torch fuel. Students were instructed to compare the brightness of the fires and the time it takes for the fires to burn out. The test was performed two times.  The first time, the same amount of fuel was used in both dishes. The second time, double or triple the amount of fuel was put into the larger dish. The students observed that the fire in the larger dish burns brighter and for less time than the fire in the smaller dish.  Even when the large dish had more fuel than the small dish, the fire in the larger dish burnt out faster.  Students then made the connection that, like the fires in the dishes, larger stars burn brighter and for less time than smaller stars. 

The analogy of burning dishes and stars is not perfect, as stars convert matter into energy via nuclear fusion reactions, while burning fuel is a chemical reaction. We found that the students did not have a problem with this difference, and appreciated just being directly told about the difference. A picture of the setup and a video of the thinking tool in action are available on the website.

\subsection{Sample Student Experience}
Students who select different questions will have varying experiences in the inquiry. While we would like to show multiple student experiences here, the space consideration limits us to one example. More examples are available on the website. We caution that these examples are merely intended to represent a wide variety of possible student experiences, and do not necessarily represent the best way to facilitate students through the inquiry. It is important to remember not to over facilitate -- it is easy to do this, and it should be avoided as much as possible. For instance, it might be okay to suggest students look at the data given, but hopefully they would come up with the idea to organize the data, and which data to organize. It is often better to ask them questions rather than telling them to do something. 

Suppose the students pick the question: Why are some stars brighter than other stars? The example below goes through what the students may be thinking at different parts of the inquiry. 

{\slshape I have lots of data, but I'm not sure what it means. I start comparing each of the spectra to the information written on them and the corresponding star in the field of stars. I start making a table as there is too much to see. Once I have a table, I'm struggling a bit to understand it. The facilitator reminds me there is graph paper, and I decide to make a graph of the different data. From the plots, I can see that brighter stars are bluer in color, and dimmer stars are redder in color. I know from the previous CLS inquiry that hot stars are blue and cool stars are red, and this is again reinforced by the plots of the data. I also see that there is a relation of the luminosity  and mass variables. The brighter bluer stars are also more massive. In fact, slight increases in mass result in large increases in the brightnesses of the stars.

I see that stars that are bluer, hotter and more massive are brighter. Where does the extra brightness come from? While trying to understand this, we are called together to go through a thinking tool. After the thinking tool, I think about the source of energy for the stars. I realize that the energy of a star depends on the mass. Specifically that mass goes as the energy, and therefore the brightness. A facilitator asked why there aren't even brighter stars seen in the field if there are stars up to 100 solar masses that exist. I estimate the lifetime of stars of different brightnesses, and see that bright stars have the shortest lives. 
 
I finally come to the conclusion that the brighter stars are stars which are the most massive, and therefore hotter stars (and also blue). The brightest stars are the most massive stars that still exist, while the fainter stars will live the longest. }

\section{Assessment}

The main evidence that the students learn what we want them to learn comes from both the facilitator and a poster presentation. The first evidence of success would be that the students could correctly answer their question, as this usually covers a lot of the content goals. In addition, we were looking for all the students to organize the data given into something that shows trends between different variables. Ideally, this would be in plots of Luminosity vs.\ Temperature, Luminosity vs.\ Mass, and Radius vs.\ Mass. If the students are experienced enough, we would see these on log-log plots. We also expect to see an understanding of how the mass of the star basically determines the evolution of a star. Ideally we would hear the students talk about how the high mass stars live less long than the low mass ones, and how the high mass, bright, and  high temperature stars are blue in color. For those studying the stars with gas around them, we would want to hear them talk about the stars being young having just formed out of the gas. For groups studying red clusters, we would want to hear about how the young stars have all died, and we only see the longer living low mass red stars, which live longer. 

In addition, since the most important goals were not content goals, but rather inquiry process goals and attitudinal goals, we expect the students to exhibit positive attitudes during the end of the inquiry and the presentations. Are the students motivated and asking lots of new questions about astronomy? Are they engaged while listening to their peers' presentations? Are the questions they ask their peers showing evidence based thinking? We expect everyone to have interpreted the data and recognized patterns. It would be an added bonus for them to be metacognitive about this, and present it as something they learned. Comments during the poster presentations  like, `Once we organized the data via graphs things just started to make sense' would be an indication that things worked. Lastly, it would be nice see students mention failed hypotheses, and how the data led them to discard them and come up with new hypotheses. 

During the investigations, the facilitators should also ask students questions to gauge their learning. For instance, once they think they have an answer to their questions, the facilitator should ask them to use their reasoning to predict answers to other questions. 

\section{Considerations for the Future}

This inquiry could be easily modified to work in a large Astronomy 101 class, especially in a classroom that uses active teaching methods. Studies in introductory astronomy classes show an increased gain in concept inventories with active teaching methods \citep{Prather:2009p9242}. It would be great to modify this inquiry to work in one of these large classroom settings, and then test the students with and without the inquiry to see how well they learned the information. It would be important to test the long term retention of the information, as inquiry has been shown to help long term retention of information \citep{hpl}. 
However as the main goal of this activity is for the students to gain process skills, the instructor should test for the development and retention of these skills as well.

\acknowledgements The authors wish to thank the Center for Adaptive Optics for making this inquiry design possible. We would especially like to thank Lisa Hunter, Lynne Raschke, Patrik Jonsson, and Hilary O'Byran for their work with the Professional Development Program where the initial design was created, as well as for giving us the opportunity to teach the inquiry in the short course at the University of California, Santa Cruz. We also thank Mark Hoffman and Elisabeth Reader for their help in teaching this inquiry in the Po`okela program at Maui Community College, in Maui, Hawaii. We thank the Akamai Workforce Initiative managed by the University of Hawaii Institute for Astronomy, Maui. This material is based upon work supported by: the National Science
Foundation (NSF) Science and Technology Center program through the Center for Adaptive Optics, managed by the University of California at Santa Cruz under cooperative agreement AST\#9876783; NSF AST\#0710699; Air Force Office of Scientific Research (via NSF AST\#0710699); University of Hawaii.

\bibliography{mrafelski}

\end{document}